\documentclass[12pt]{iopart}
\usepackage{times,xspace}
\usepackage{amsbsy,amssymb,bm,mathptmx}
\usepackage{graphicx,color,epsfig,rotate}
\usepackage{fancyhdr}

\begin{document}

\article[Electromagnons in multiferroic $R$Mn$_2$O$_5$]{Focus Issue
Article}{Electromagnons in multiferroic $R$Mn$_2$O$_5$ compounds and their microscopic origin}

\author{A B Sushkov$^{1}$, M Mostovoy$^{2}$, R Vald\'{e}s Aguilar$^{1}$, S-W Cheong$^{3}$ and H D Drew$^{1}$}

\address{$^{1}$ Materials Research Science and Engineering Center, University of Maryland, College Park, Maryland 20742}
\address{$^{2}$ Zernike Institute for Advanced Materials,
University of Groningen, 9747 AG Groningen, The Netherlands}
\address{$^{3}$ Rutgers Center for Emergent materials and Department of Physics and Astronomy, Rutgers University, Piscataway, New Jersey 08854}
\ead{sushkov@umd.edu}
\date{\today}
\begin{abstract}
We summarize the existing experimental data on electromagnons in 
multiferroic $R$Mn$_2$O$_5$ compounds, where $R$ denotes a rare earth 
ion, Y or Bi, and discuss a realistic microscopic model of these 
materials based on assumption that the microscopic mechanism of   
magnetically-induced ferroelectricity and electromagnon absorption 
relies entirely on the isotropic Heisenberg exchange and magnetostrictive 
coupling of spins to a polar lattice mode 
and does not involve relativistic effects. This model 
explains many magnetic and optical properties of $R$Mn$_2$O$_5$ manganites, such 
as the spin re-orientation transition, magnetically-induced 
polarisation, appearance of the electromagnon peak in the non-collinear 
spin state and the polarisation of light for which this peak is 
observed.  We compare experimental and theoretical results on 
electromagnons in $R$Mn$_2$O$_5$ and $R$MnO$_3$ compounds.
\end{abstract}

\maketitle

\section{Introduction}
\label{sec:introduction}

Multiferroic materials that exhibit simultaneous magnetic and ferroelectric order have attracted much attention recently because of the fundamental interest of systems with coupled order parameters and because of their potential for cross electric and magnetic functionality
\cite{Smol-Chupis,Fiebig2005,Prellier2005,Khomskii2006,Eerenstein2006,Cheong2007,Ramesh2007}.  Many recently discovered multiferroics, e.g. TbMnO$_3$ \cite{Kimura-113}, TbMn$_2$O$_5$ \cite{Hur-nature}, and Ni$_3$V$_2$O$_8$ \cite{Lawes-vanadate}, are improper ferroelectrics in which the electric polarisation is induced by spin ordering. These materials show striking cross-coupling effects such as magnetic field
induced polarisation switching \cite{Kimura-113,Hur-nature} and giant magnetocapacitance~\cite{Goto2004} and a strong coupling between spin and lattice excitations that leads to electric dipole excitation of magnons, or electromagnons~\cite{Smol-Chupis,Pimenov-Nature,Sushkov125,Katsura-em}, which is the subject of this paper. The fundamental interest in multiferrocity also derives from the strong interplay between magnetic frustration, ferroelectric order and unusual symmetry breaking in phase transformations that characterize these materials \cite{Cheong2007,Kimura07}.

The known microscopic mechanisms of magnetically-induced
ferroelectricity include lattice distortion (exchange striction) and redistribution of electron density in response to spin ordering. Such processes occur locally in all magnetic materials. 
However, only when a spin ordering breaks inversion symmetry do these local electric dipoles add into a macroscopic electric polarisation. 
Spin orders that break inversion symmetry are rare and the best systems to look for them are frustrated magnets, where competing interactions and the geometry of spin lattice favor unconventional magnetic states. In most of the recently discovered multiferroic materials, such as TbMnO$_3$, Ni$_3$V$_2$O$_8$, MnWO$_4$ and CuO, competing interactions force spins to form a cycloidal spiral. This non-collinear spin order breaks inversion symmetry and activates antisymmetric Dzyaloshinskii-Moriya interaction proportional to the cross-product of spins, $\bi{S}_{1} \times \bi{S}_{2}$ \cite{Moriya}. The concomitant lattice and electronic distortion induces electric polarisation \cite{Katsura2005,SergienkoDagotto,Jia}. 

When inversion symmetry is broken by a collinear magnetic ordering the strongest spin interaction that can
shift ions and polarise electronic clouds is the symmetric Heisenberg exchange, proportional to the scalar product of
spins, $\bi{S}_{1} \cdot \bi{S}_{2}$. This mechanism  was proposed to explain multiferroic properties of $R$Mn$_2$O$_5$, where $R$ denotes a rare earth ion, Y or Bi, and orthorombic manganites showing the E-type antiferromagnetic ordering \cite{Kadomtseva,ChaponPRL2006,SergienkoPRL2006}.

These two microscopic mechanisms of magnetically-induced ferroelectricity give rise to two different forms of phenomenological magnetoelectric coupling:  electric polarisation induced by a low-pitch spiral is described by the third-order coupling term $P \left(L_{1} \partial L_{2} - L_{2} \partial L_{1}\right)$, where $P$ is electric polarisation and $L_{1,2}$ are magnetic order parameters describing the sinusoidal and cosinusoidal components of the spiral \cite{Baryachtar,Mostovoy-spiral}, while the coupling working in collinear spin states has the form $P\left(L_{1}^2-L_{2}^2\right)$, where $L_{1}$ and $L_{2}$ are components of a two-dimensional irreducible representation describing magnetic states with opposite electric polarisations \cite{Levanyuk,Kadomtseva,SergienkoPRL2006}.

\begin{figure}
\centering
\includegraphics[width=8cm]{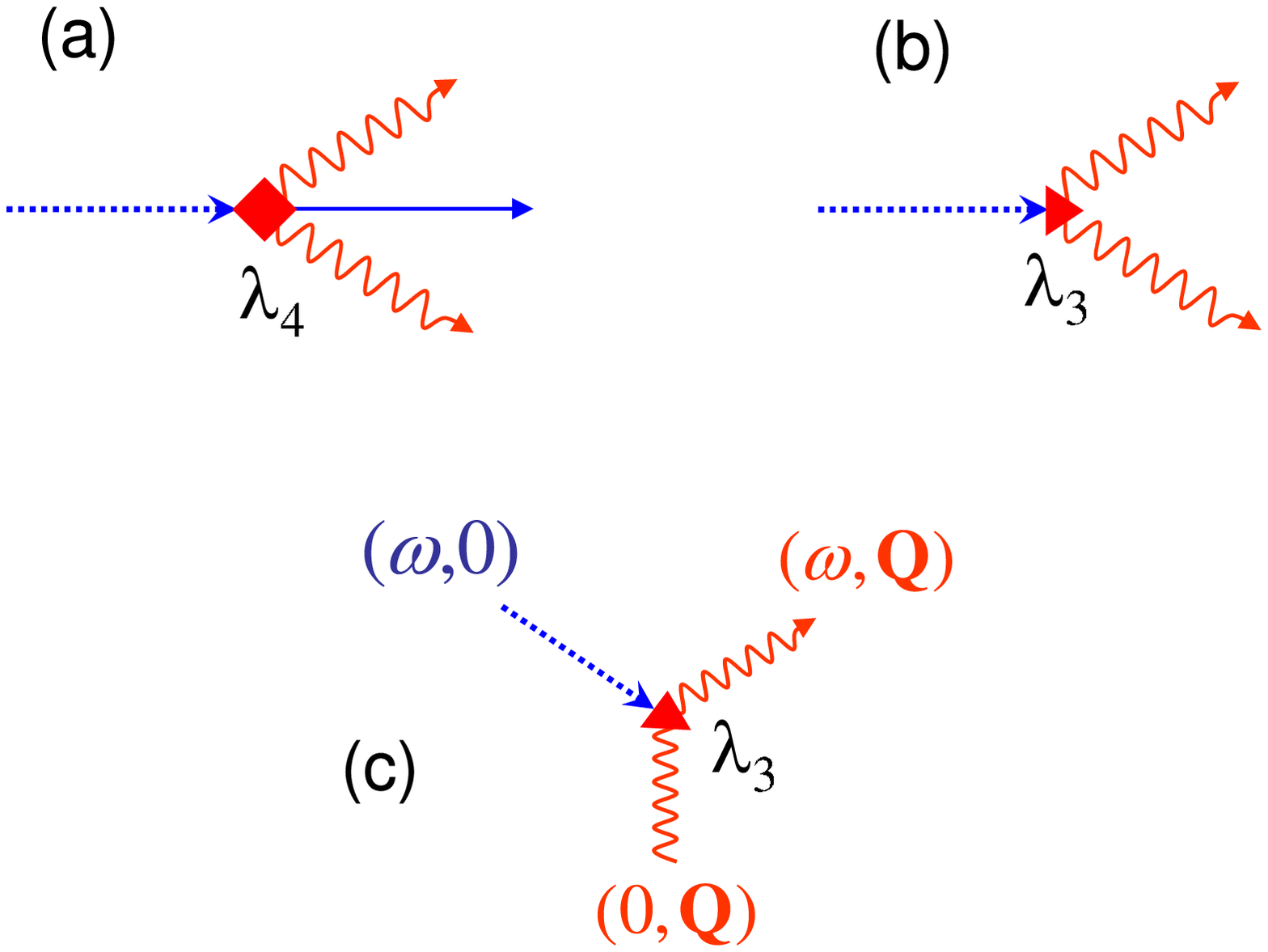}
\caption{(Colour online) Feynman diagrams describing photo-excitation of magnons by the electric field of light (here photon and phonon are represented, respectively, by dashed and solid lines, while the wavy line corresponds to magnon): (a) Photoexcitation of two magnons and one phonon via the fourth-order magnetoelectric coupling (the Lorenzana-Sawatzky mechanism), (b) Excitation of the two-magnon continuum via the third-order magnetoelectric coupling (`charged magnons') and (c) Photoexcitation of a single magnon (electromagnon) due to the third-order magnetoelectric coupling, where photon with the frequency $\omega$
and zero wave vector scatters off the static spin modulation with the wave vector ${\bf Q}$ producing a magnon with the same wave vector and the frequency ${\omega}$.} 
\label{fig:photoex}
\end{figure}

The magnetoelectric interactions that induce electric polarisation in magnets can also couple oscillations of magnetisation to polar lattice vibrations. The oscillations of polarisation at magnon frequency and vice versa give rise to dynamic magnetoelectric effects, such as electromagnon excitations. In usual magnets, an oscillating electric field of photons can excite a three-particle continuum consisting of two magnons and one phonon \cite{Lorenzana1995}. This process results from the fourth-order spin-lattice coupling (see figure~\ref{fig:photoex}a). The third-order couplings in multiferroics, discussed above, make possible photo-excitation of two-magnon continuum without a phonon (`charged magnons' \cite{Damascelli}) shown in figure~\ref{fig:photoex}b). Replacing one of the magnons by the static modulation of spin density appearing in the ordered spin state, we obtain a process that converts photon into a single magnon, which is the electromagnon (see figure~\ref{fig:photoex}c). This process is usually mediated by a polar phonon linearly coupled to both magnons and light, which for low-frequency phonons can lead to a resonant enhancement of the photo-excitation of electromagnons.     

As there are two possible contributions to the polarisation --- from ionic displacements and from electronic density redistribution --- one can think of two corresponding electric contributions to electromagnon --- from phonons and from electronic excitations. 
This means a transfer of the electric dipole spectral weight from phonons ($\hbar\omega \sim 10 - 100$~meV) and/or electronic excitations
($\hbar\omega \sim 2$~eV) down to magnons ($\hbar\omega \sim 1 - 5$~meV). Such a transfer, in turn, leads to a step-like anomaly in the temperature dependence of the dielectric constant. Current research on electromagnons, including this paper, is focused mainly on magnon-phonon coupling while magnon-electron aspect is much less explored as it is generally expected to be weak because of the large energy difference between the electronic excitations and the magnetic excitations.

The possibility of electromagnon excitations in multiferroics has long been anticipated theoretically~\cite{Smol-Chupis}, but only
recently were they observed in experiment. 
Pimenov \etal~\cite{Pimenov-Nature} reported observation of electric active modes at magnon frequencies in GdMnO$_3$ and TbMnO$_3$ which exist only in some magnetically ordered phases and can be suppressed by magnetic field. 
Further, they found in GdMnO$_3$ the spectral weight transfer from the lowest frequency phonon down to the electromagnon mode~\cite{Pimenov-PRB}. 
Sushkov \etal~\cite{Sushkov125} reported observation of electromagnons in YMn$_2$O$_5$ and TbMn$_2$O$_5$.
Electromagnons in both these $R$Mn$_2$O$_5$ compounds have very similar spectrum and exist only in magnetic ferroelectric phases that proves their magnetic origin. 
Vald\'{e}s Aguilar \etal~\cite{Valdes-EuY} showed that electromagnon absorption in Eu$_{0.75}$Y$_{0.25}$MnO$_3$
 occurs over a broad band with two peaks, both of which exist only in the magnetic ferroelectric phase below 30~K. 
 Pimenov \etal~\cite{Pimenov-EuY} explored a composition set Eu$_{1-x}$Y$_{x}$MnO$_3$ for $0 \leq x \leq 0.5$ and confirmed main features of electromagnons in $R$MnO$_3$. Kida~\etal~\cite{Kida2008} found out that the electromagnon polarisation in DyMnO$_3$ stays with the lattice when the spin plane is rotated by the external magnetic field. 
Electromagnons, observed so far only in non-collinear spin phases of $R$MnO$_3$ and $R$Mn$_2$O$_5$ compounds, have common features for both families: they are active only in one polarisation $\bi{e} \| \bi{a}$ axis for $R$MnO$_3$ and $\bi{e} \| \bi{b}$ axis for $R$Mn$_2$O$_5$, where $\bi{e}$ is electric field of light; well defined peaks exist only in the low-temperature magnetic ferroelectric phase. 
A theory of electromagnons for the circular magnetic spiral was developed by Katsura~\etal~\cite{Katsura-em} and its continuum version was given by de Sousa \etal~\cite{deSoussaPRB2008,deSousa} who applied it to describe low-energy excitations in BiFeO$_3$. Fang and Hu~\cite{FangCM} discussed electromagnon absorption in $R$Mn$_2$O$_5$ compounds assuming Heisenberg interactions between spins. 

The outstanding fundamental questions for electromagnon are the 
microscopic origin of these novel excitations (Heisenberg or Dzyaloshinskii-Moriya type exchange) in the 
different classes of compounds ($R$MnO$_3$ and $R$Mn$_2$O$_5$), the explanation of the 
observed selection rules, and whether electromagnons may occur in a 
wider range of materials.  The practical issues are enhancing the 
magneto-capacitance effect and its temperature range and possibly 
applying these new excitations to metamaterials and/or achieving 
negative index of refraction in the magnon range of frequencies.

The remainder of this paper is organized as follows. In section~\ref{sec:exp125}, we discuss optical absorption spectra of multiferroic YMn$_2$O$_5$ and TbMn$_2$O$_5$ compounds taken at different temperatures. We present experimental evidence allowing us to identify the low-frequency peaks appearing in the non-collinear phase of these materials as electromagnons. For comparison, in section~\ref{sec:exp113} we present data on electromagnons peaks in the spiral multiferroic material Eu$_{0.75}$Y$_{0.25}$MnO$_3$. In section~\ref{sec:model}, we discuss a simple microscopic model of YMn$_2$O$_5$ manganites, based on isotropic Heisenberg exchange interactions between spins, with which we describe magnetic, multiferroic and optical properties of these compounds. We also briefly discuss phenomenological description of magnetic orders and magnetoelectric coupling in $R$Mn$_2$O$_5$ compounds based on their symmetry. In section~\ref{sec:discussion}, we discuss the microscopic origin of the electromagnon peaks in both $R$Mn$_2$O$_5$ and $R$MnO$_3$ and how the electromagnon relates to the spontaneous polarisation. Finally, we summarize our experimental and theoretical results.   

\section{Far infrared spectroscopy of $R$Mn$_2$O$_5$ compounds}
\label{sec:exp125}

\begin{figure}
\centering
\includegraphics[width=8cm]{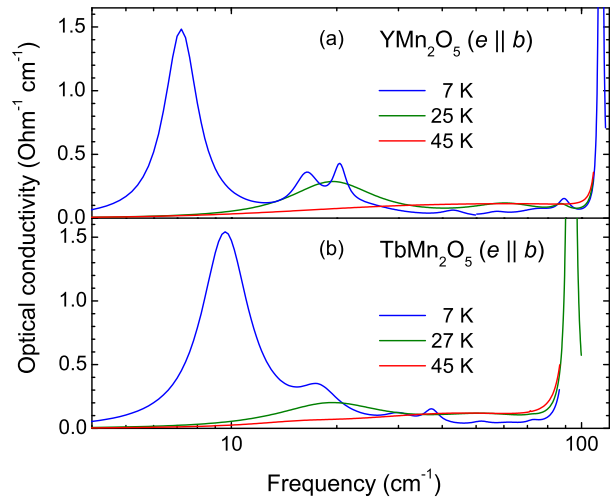}
\caption{(Colour online) Optical conductivity of YMn$_2$O$_5$ and TbMn$_2$O$_5$ for the electric field of light $\bi{e} \| \bi{b}$ in three phases. Strong peaks at 113 and 97~cm$^{-1}$ are the lowest phonons, other peaks are electromagnons.} \label{s1}
\end{figure}

The family of $R$Mn$_2$O$_5$ compounds has long been studied~\cite{Bertaut1967}. 
Small variations of the exchange integrals with temperature lead to the complex phase diagram for this multiferroic family~\cite{Harris2008}.
Earlier spectroscopic works revealed far infrared absorption modes activated at low temperatures in 
EuMn$_2$O$_5$ \cite{Sanina1988}, YMn$_2$O$_5$ \cite{Mukhin2000}  and  GdMn$_2$O$_5$ \cite{Golovenchits2003}. 
Recently, we have shown that in YMn$_2$O$_5$ and TbMn$_2$O$_5$ strong electric dipole active modes emerge at magnon frequencies in the lowest in temperature ferroelectric phase~\cite{Sushkov125}. 
As in the work by Pimenov \etal~\cite{Pimenov-Nature} on $R$MnO$_3$ compounds, we assign this modes as electromagnons.   

In figure~\ref{s1}, we show the optical conductivity spectra of YMn$_2$O$_5$ and TbMn$_2$O$_5$ at three different temperatures corresponding to three magnetic/ferroelectric phases. These spectra were obtained by fitting measured transmission spectra~\cite{Sushkov125} with the model dielectric function~\cite{reffit}, assuming all modes to be of the electric dipole nature. The spectra measured at 7~K (blue curves) show the  characteristic strong and sharp (electromagnon) absorption peaks at lowest frequencies. In this phase, the spontaneous electric polarisation is relatively small and the angles between  neighbouring spins along the $b$ axis are large. The spectra measured at $25$~K (green curves) have one broad absorption peak near 20~cm$^{-1}$. In this phase spins are almost collinear and electric polarisation is large. Red curves, taken just above the N\'{e}el temperature, show a single broad absorption band below the phonon frequencies.

Identifying the low-frequency excitations as electromagnons requires addressing several questions. First, to avoid confusion with possible transitions between $f$-levels of rare earth ions, we have studied YMn$_2$O$_5$. The second issue is electric versus magnetic dipole (antiferromagnetic resonance) activity. 
We measured transmission spectra for various mutual orientations of the electric and magnetic field of light (respectively, $\bi{e}$ and $\bi{h}$) with respect to the crystal axes and we found the absorption only for $\bi{e} \| \bi{b} \| \bi{P}$, where $\bi{P}$ is the spontaneous polarisation vector, independent of the orientation of $\bi{h}$, which implies that the excitations are electric dipole active. Can these resonances be new phonons, activated in the low
temperature phase? We have performed shell model calculations that put the lowest phonon near 100~cm$^{-1}$ for any reasonable parameters. Thus, we believe that we can reliably identify these low-frequency peaks as electromagnons.

Another check of the electromagnon origin of this modes is comparison of their contribution to the step-like anomaly in the temperature dependence of $\varepsilon_{1}$, calculated using the Kramers-Kronig relation, with the measured one.
\begin{figure}
\centering
\includegraphics[width=8cm]{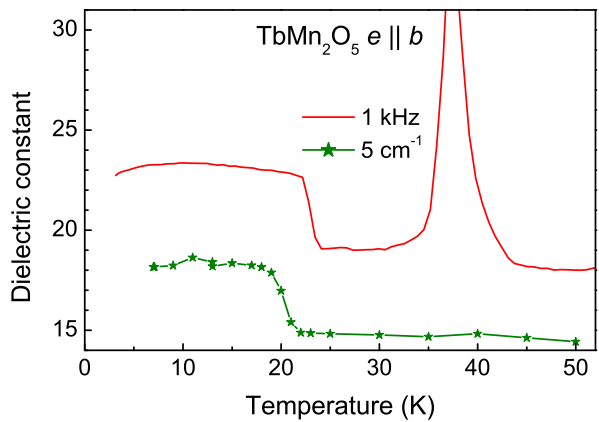}
\caption{(Colour online) Dielectric constant of TbMn$_2$O$_5$ from fits of infrared spectra (lower curve) in comparison with kHz measurements. The whole step-like anomaly is due to electromagnons (figure~\ref{s1}a, blue curve).}
\label{figeps}
\end{figure}
Figure~\ref{figeps} shows such a comparison. We have chosen  TbMn$_2$O$_5$ for this purpose because of the larger sample size and higher frequency of the electromagnon peaks. It is clear from figure~\ref{figeps} that the whole step-like anomaly in $\varepsilon_1(T)$ comes entirely from the sharp electromagnon peak.

The frequency and temperature behaviour, presented in figures~\ref{s1} and \ref{figeps} is typical for a set of $R$Mn$_2$O$_5$ compounds: 
$R$ = Er, Ho, Y, Dy, Tb, Gd and Eu~\cite{Kobayashi2004,Sirenko-Ho,Sushkov125,Golovenchits2003,Sanina1988,RolandoPhD}. 
The appearance of electromagnon peaks as well as the $\varepsilon$ step-like anomaly seem to be correlated with the transition into the non-collinear spin state. 
Notably, BiMn$_2$O$_5$ whose spin ordering is nearly collinear at all temperatures, shows neither the $\varepsilon$ step-like anomaly nor electromagnon absorption~\cite{RolandoPhD}.

An inelastic neutron scattering study of YMn$_2$O$_5$ is in progress. 
Preliminary data by S.-H.~Lee \etal~\cite{LeeAPS} show several scattering peaks in energy scans at the wave vector of the static spin structure. 
A strong neutron feature is observed at 1~meV in good agreement with the sharp low frequency feature in the low temperature infrared spectrum (figure~\ref{s1}a). 

Katsura \etal~\cite{Katsura-em} predict that the electromagnon originating from antisymmetric exchange has a `transversal' polarisation with respect to the spontaneous polarisation: $\bi{e} \perp \bi{P}$.
The observed polarisation selection rule for electromagnons in $R$Mn$_2$O$_5$ compounds is `longitudinal': $\bi{e} \| \bi{P}$. 
We will show that such `longitudinal' electromagnon can be obtained in a model based on the isotropic Heisenberg exchange. 

\section{Far infrared spectroscopy of $R$MnO$_3$}
\label{sec:exp113}

It is interesting to compare electromagnons in the two families of multiferroic manganites. We begin with Eu$_{0.75}$Y$_{0.25}$MnO$_3$ --- a compound that mimics lattice parameters of TbMnO$_3$ but does not have $f$--$f$ transitions in the far infrared~\cite{Valdes-EuY}.
\begin{figure}
\centering
\includegraphics[width=8cm]{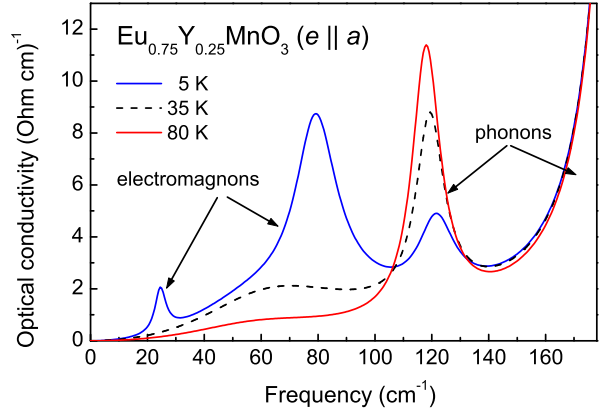}
\caption{(Colour online) Optical conductivity of
Eu$_{0.75}$Y$_{0.25}$MnO$_3$ for $\bi{e} \| \bi{a}$ in the
ferroelectric (blue), the spin density wave (dashed), and the paramagnetic (red) phases. Electromagnon band consists of a broad background and two peaks. } \label{s1EuY}
\end{figure}
To extract the parameters of the oscillators, we fit the transmission spectra 
with a Lorentzian model of the dielectric constant $\varepsilon(\omega)$ for electric dipole transitions:
\begin{equation}
	\varepsilon(\omega) = \varepsilon_\infty+\sum_j
\frac{S_j}{\omega_{j}^2-\omega^2-\imath\omega\gamma_j}
\label{Lorentz}
\end{equation}
where $\varepsilon_\infty $ is the high frequency dielectric constant, $j$ enumerates the oscillators,
$S_j$ is the spectral weight, $\omega_{j}$ is the resonance frequency, and $\gamma_j$ is the damping rate. 

\begin{figure}
\centering
\includegraphics[width=8cm]{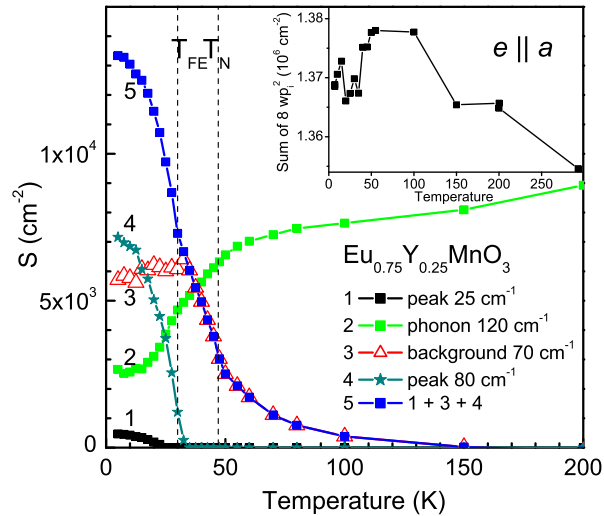}
\caption{(Colour online) Spectral weight of the absorption peaks below 140~cm$^{-1}$ in figure~\ref{s1EuY}. 
The numbers from 1 to 5 enumerate the curves. Frequencies in the legend are the lowest temperature resonance frequencies of each mode. 
Inset: Total spectral weight of the eight phonons above 140~cm$^{-1}$.} \label{swEuY}
\end{figure}

Figure~\ref{s1EuY} shows optical conductivity spectra for three phases of Eu$_{0.75}$Y$_{0.25}$MnO$_3$. 
Figure~\ref{swEuY} shows temperature dependence of the fit parameter $S_j$ (spectral weight of the peaks in figure~\ref{s1EuY}).
A broad absorption (electromagnon) band exists well above the N\'{e}el temperature 47~K (red curve in figure~\ref{s1EuY} and curve~3 in figure~\ref{swEuY}). 
Upon cooling, the spectral weight $S$ of this broad band is growing down to $T_{FE}$=30~K and stays at this level at lower temperatures. 
Such a behaviour of the background can be seen from the spectra --- the absorption at 40~cm$^{-1}$, the frequency least affected by two absorption peaks, stays constant at all $T < T_{FE}$. 
This absorption produces smooth growth of the $\varepsilon_a(T)$ at $T_{FE} < T$. $T_N$=47~K is the inflection point of the curve~3 which is an
evidence for the magnetic origin of this low frequency broad electric absorption. 
Two electromagnon peaks emerge sharply at the $T_{FE}$ (blue curve in figure~\ref{s1EuY} and curves 1 and 4 in figure~\ref{swEuY}). They produce all further growth of the $\varepsilon_a(T)$ at $T < T_{FE}$. The electromagnon spectrum of TbMnO$_3$ is very similar to that one of Eu$_{0.75}$Y$_{0.25}$MnO$_3$, except
the large electromagnon peak is at 60 instead of 80~cm$^{-1}$ \cite{{RolandoPhD}}.

It is also interesting to follow the temperature dependence of the redistribution of the electric dipole spectral weight $S$. Since electromagnons result from a small admixture of phonons to magnons, the total spectral weight should be conserved. Comparing the curves 2 and 5 in figure~\ref{swEuY}, one can estimate how much spectral weight the 120~cm$^{-1}$ phonon is loosing and how much electromagnons acquire. As electromagnons gain more spectral weight than the phonon is loosing, we checked the
rest of the phonon peaks for the same polarisation of light. The inset in figure~\ref{swEuY} shows that, indeed, phonons lose just enough of the spectral weight to conserve total spectral weight.

In their recent inelastic neutron scattering work, Senff \etal~\cite{SenffPRL} reported a set of modes at the incommensurate zone center of TbMnO$_3$. 
The frequency of the lowest mode is equal to the frequency of the lowest infrared peak (24~cm$^{-1}$). 
We agree with Senff~\etal in assignment of this mode as electromagnon. 
However, despite the satisfied polarisation selection rule $\bi{e} \perp \bi{P}$, this is not the electromagnon predicted by Katsura \etal~\cite{Katsura-em}, as we discuss below.   

\section{The model of $R$Mn$_2$O$_5$}
\label{sec:model}

The presence of two different types of magnetic ions and geometric frustration of spin interactions give rise to rather complex magnetic structures $R$Mn$_2$O$_5$ compounds. Nonetheless, a number of salient properties of these materials, such as the magnetically-induced electric polarisation, photo-excitation of magnons as well as the spin re-orientation transition, can be understood within a simplified microscopic model, which we discuss in this section. Our starting point is the assumption that multiferroic and optical properties of these materials are governed by the isotropic Heisenberg exchange, although we do include magnetic anisotropy to explain the spin  re-orientation transition that has a strong effect on the low-frequency absorption spectrum. We discuss ordered spin states of the model, the mechanism of magnetoelectric coupling and calculate the optical absorption spectrum at magnon frequencies for different magnetic states.

We start by considering a single magnetic $ab$-layer including Mn$^{3+}$ and Mn$^{4+}$ ions. The model describes interactions between the spins and their coupling to a polar phonon mode:
\begin{equation}
\begin{array}{ll}
H = & \frac{1}{2}\sum_{i,j}J_{ij}\left( {\bf P}\right) \left( \bi{S}_{i}
\cdot \bi{S}_{j}\right) -\frac{1}{2}\sum_{i\alpha } K_{i\alpha } \left( \bi{S}_{i} \cdot \hat{k}_{i\alpha }\right)^{2}- \\
& -\sum_{i}\mu _{i}\left( \bi{S}_{i} \cdot \bi{H}\right) +V\left(\frac{\bi{P}^{2}}{2\chi _{1}^{(0)}}-\bi{PE} - \frac{\chi_{2}
E^{2}}{2} \right).
\end{array}
\label{eq:phi}
\end{equation}
Here the first term is the isotropic Heisenberg spin exchange, while the second term is the single-ion anisotropy. The antisymmetric Dzyaloshinskii-Moriya exchange as well as other types of anisotropic exchange interactions are not included, as in the scenario discussed below they play no role. We assume for simplicity that in the ordered state all spins lie in the $ab$ plane
\cite{ChaponPRL2006} and neglect the small out-of-plane components found in recent neutron diffraction experiments on single crystals \cite{Noda2006,Vecchini2008}. Thus, the easy and intermediate magnetic axes on each Mn site ($\alpha = 1,2$) lie in the $ab$ plane, while the hard axis $\hat{k}_{i3}\parallel \hat{c}$. The third term in (\ref{eq:phi}) is the interaction of spins with an applied magnetic field and the last term describes the dielectric response of the system, where $\chi _{1}^{(0)}$ is the `bare' dielectric susceptibility related to the polar lattice mode (not including the magnetic contribution calculated below) and $\chi _{2}$ is the remaining part of the dielectric susceptibility of non-magnetic origin. Finally, $V$ is the system volume.

The coupling between the spins and the polar phonon mode results from the dependence of the exchange coupling on the electric polarisation, which in $R$Mn$_2$O$_5$ is parallel to the $b$ axis:
\begin{equation}
J_{ij}(P_{b}) = J_{ij}(0) + J_{ij}^{\prime }(0)P_{b} + \frac{1}{2}
J_{ij}^{\prime\prime }(0)P_{b}^2 + \ldots  \label{eq:J(P)}
\end{equation}
The last two terms give rise to the cubic and the quartic  magneto-electric couplings.

\subsection{Magnetic ordering and spin re-orientation transition}

\begin{figure}[tbh]
\begin{center}
\includegraphics[width=8cm]{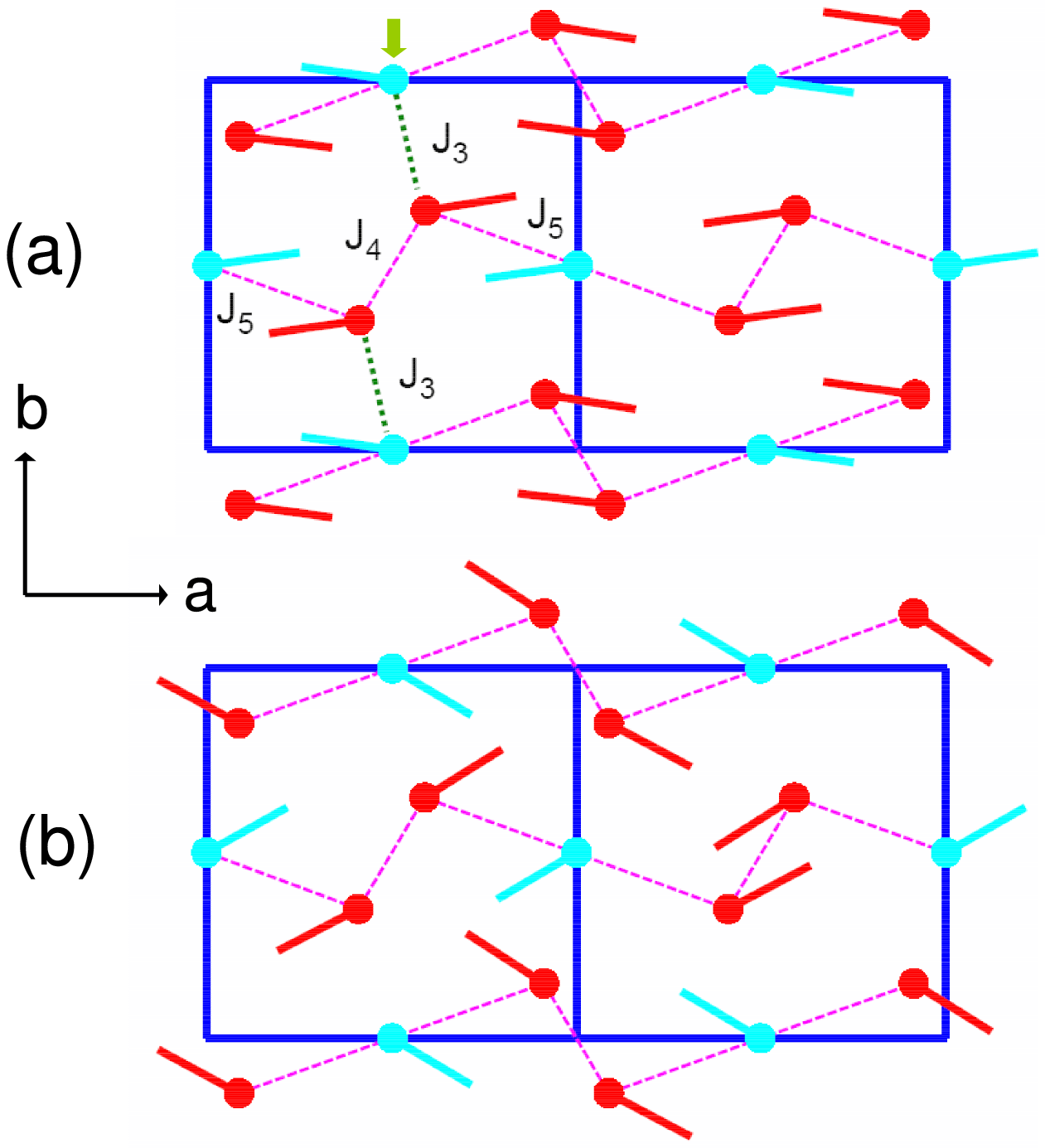} \vspace{0.0cm}
\caption{(Colour online) Minimal energy spin
configurations for $J_{4}=J_{5}=40$~K and the anisotropy parameter $K_{a}(\mbox{Mn}^{3+})=0.6$~K  for all Mn$^{3+}$ ions  (red) and $K_{a}(\mbox{Mn}^{4+}) =0.1$~K for all Mn$^{4+}$ ions (blue). The value of the interchain coupling $J_{3}$ is $-2$~K for the structure in panel (a) and $-4$~K for the one panel (b).}
\label{fig:ordering}
\end{center}
\end{figure}

Here, we adopt the model of Chapon~\etal~\cite{ChaponPRL2006} with 5 exchange constants between pairs of nearest-neighbour Mn ions:
$J_{1}$ and $J_{2}$ couple Mn$^{4+}$ ions along the $c$ direction, $J_{3}$ and $J_{5}$ couple the spins of neighbouring Mn$^{3+}$ and Mn$^{4+}$ ions and  $J_{4}$ is the coupling between two neighbouring Mn$^{3+}$ ions (see figure \ref{fig:ordering}).

Figure~\ref{fig:ordering} shows the minimal energy spin configurations obtained by the numerical minimization of the spin energy (\ref{eq:phi}) on the subspace of the commensurate magnetic states with the wave vector
$\bi{Q} = \left(1/2,0,0\right)$ for two different sets of exchange constants. The exchange constants $J_{4}$ and $J_{5}$ were chosen to be positive and large compared to other exchange constants, which gives rise to antiferromagnetic zig-zag chains along the $a$ axis with nearly collinear spins (marked by dashed lines) observed in neutron experiments \cite{ChaponPRL2006,Vecchini2008}.

The angle between spins in neighbouring chains sensitively depends on the ratio between the interchain coupling $J_{3}$ and the magnetic anisotropy parameters $\bi{K}_{i}$. We assume that the easy magnetic axis is parallel to the $a$ axis. The interplay between magnetic anisotropy and interchain interaction determines the angle between spins in neighbouring antiferromagnetic chains.  

We first note that if spins in each antiferromagnetic chain would be perfectly collinear than the interchain interactions would cancel as a result of geometric frustration. Conversely, the interchain coupling $J_{3}$ results in spin rotations, which destroy the collinearity of spins in each chain. Consider, for example, the spin of the Mn$^{4+}$ ion, marked in figure~\ref{fig:ordering}(a) by an arrow. Because of a nonzero angle between spins in the neighbouring $a$-chains, the interaction of the marked spin with the spin of the Mn$^{3+}$ ion from the neighbouring chain will give rise to a rotation of the marked spin. These small spin rotations lift the frustration and lead to some energy gain due to interchain interactions. It easy to show that this energy gain is maximal when spins in neighbouring antiferromagnetic chains are orthogonal to each other. Thus, while the magnetic anisotropy favours an almost collinear spin configuration, interchain interactions favour the $90^{\circ}$ angle between spins of neighbouring chains.

This competition gives rise to a very strong sensitivity of the angle between spins in neighbouring chains to the interchain coupling $J_{3}$. For weak interchain coupling $J_{3} = -2$~K, this angle is relatively small and minimal-energy spin configuration shown in figure~\ref{fig:ordering}(a) is similar to the one observed in the high-temperature `collinear' phase of YMn$_2$O$_5$ by Chapon \etal \cite{ChaponPRL2006}. A small change in $J_{3}$ from $-2$~K to $-4$~K transforms the configuration shown in figure~\ref{fig:ordering}(a) into the one shown in figure~\ref{fig:ordering}(b), which may explain the spin re-orientation observed in $R$Mn$_2$O$_5$ with $R = $Tb, Ho, Dy and Y, provided that the interchain coupling is temperature-dependent. Although the rotations that make spins in each chain non-collinear are barely visible, they are sufficient to produce the large changes in the spin configuration, since the magnetic anisotropy is relatively weak. In $R$Mn$_2$O$_5$, the spin re-orientation transition is accompanied by the loss of commensurability of the spin structure in the $a$ and $c$ directions (which is also a consequence of magnetic frustration). This latter aspect of the transition is, however, less important for the photo-excitation of magnons, discussed below, than the re-orientation of spins.

\subsection{Magnetically-induced polarisation}

Minimizing (\ref{eq:phi}) with respect to $P_{y}$, we obtain expression for the magnetically-induced electric polarisation,
\begin{equation}
P_{b} \approx -\frac{\chi _{1}^{(0)}}{2V}\sum_{i,j}J_{ij}^{\prime }\left(
0\right) \left( \bi{S}_{i}\bi{S}_{j}\right),   \label{eq:P0}
\end{equation}
which only involves scalar products of spins.

\begin{figure}[tbh]
\begin{center}
\includegraphics[width=8cm]{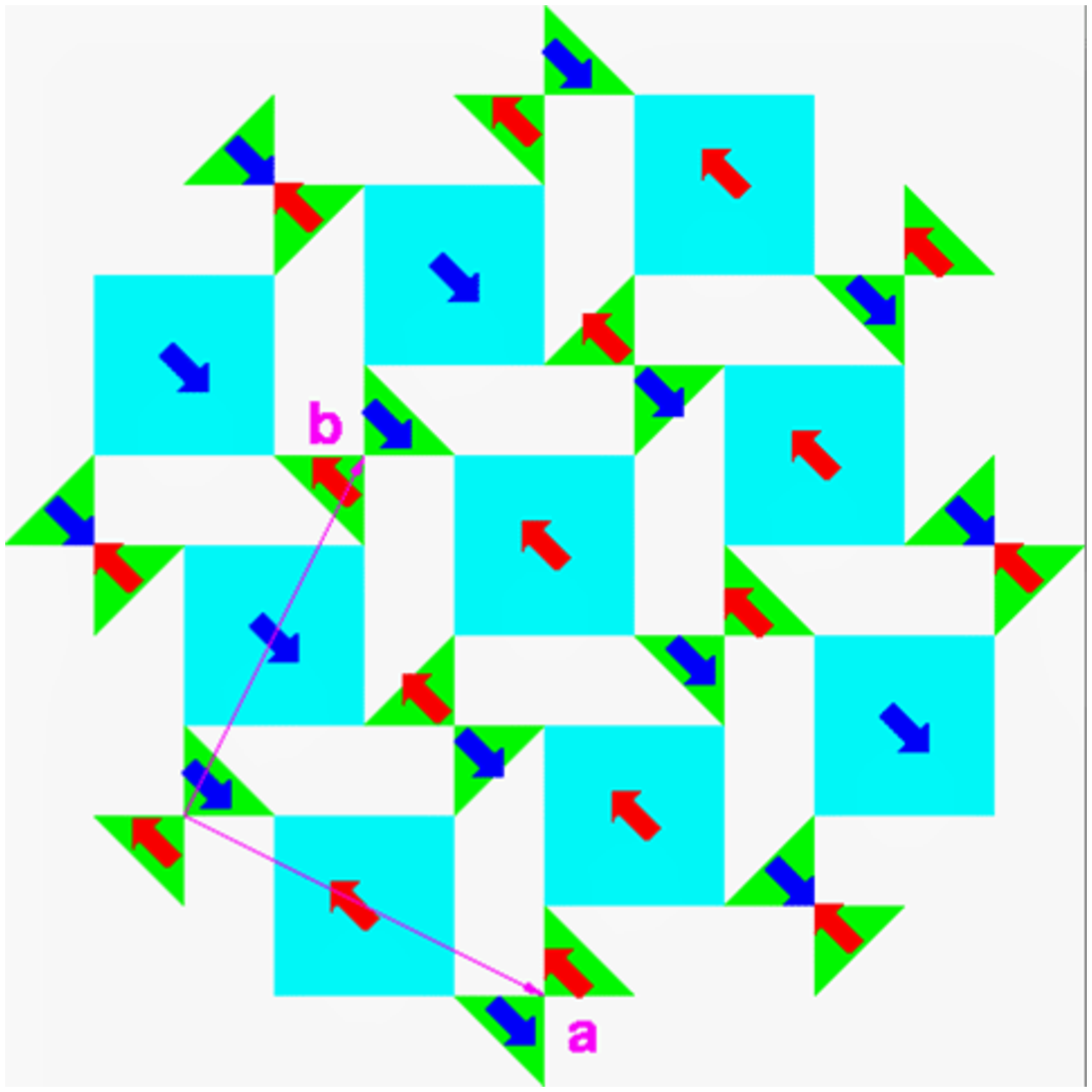} \vspace{0.0cm}
\caption{A cartoon of the magnetic ordering in the high-temperature collinear phase.}
\label{fig:cartoon}
\end{center}
\end{figure}

\begin{figure}[tbh]
\begin{center}
\includegraphics[width=12cm]{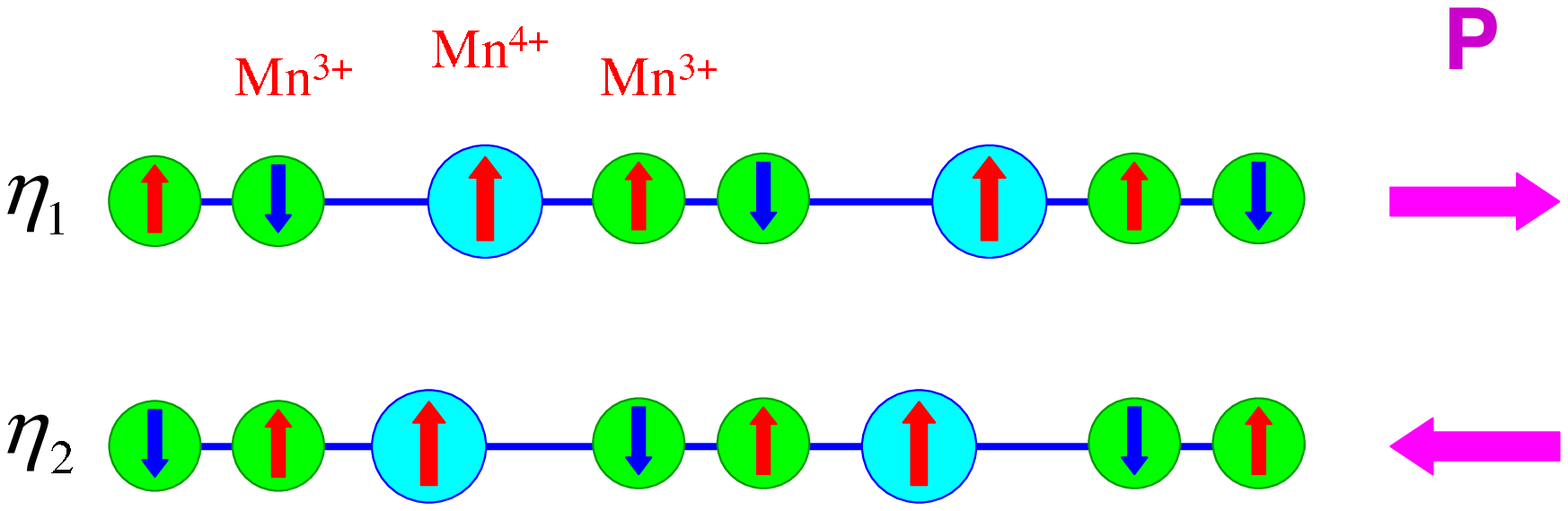} \vspace{0.0cm}
\caption{Electric polarisation induced by magnetostriction along the $b$-chains. The two order parameters $\eta_{1}$ and $\eta_{2}$ describe degenerate ferroelectric states with opposite directions of electric polarisation.}
\label{fig:dipole}
\end{center}
\end{figure}

Figures~\ref{fig:cartoon} and \ref{fig:dipole} show why in the high-temperature ferroelectric phase the polarisation vector is oriented along the $b$ axis. Figure~\ref{fig:cartoon} gives a simplified view of the Mn layer in the $ab$ plane, in which the spins inside squares depict the spins of Mn$^{4+}$ ions located inside oxygen octahedra, while the spins inside triangles
are the spins of Mn$^{3+}$ ions in oxygen pyramids. The nearly collinear magnetic ordering in the high-temperature ferroelectric phase consists of antiferromagnetic chains along the $a$ direction.

The mechanism responsible for electric polarisation in this magnetic state involves, however, the $\uparrow\uparrow\downarrow$ and $\downarrow\downarrow\uparrow$ spin chains along the $b$ axis, such as shown in figure~\ref{fig:dipole}. These chains contain the polar Mn$^{4+}$ -- Mn$^{3+}$ bonds, connecting parallel spins, and the Mn$^{3+}$ -- Mn$^{4+}$ bonds with opposite polarity, connecting antiparallel spins. Importantly, the charge and spin modulations in the chains have the same period, in which case the conventional exchange striction destroys the cancellation of electric dipoles of the polar bonds and induces electric polarisation along the chains, as illustrated in figure~\ref{fig:dipole}.

This mechanism also works in the low-temperature (incommensurate) ferroelectric phase, which has the same periodicity in the $b$-direction. The amplitude of the exchange striction is, however, the largest for collinear spins [see (\ref{eq:P0})], which explains the drop of the
polarisation at the transition to the low-temperature phase. For example, if the value of the magnetoelectric coupling $\propto J_{3}^{\prime} - J_{4}^{\prime} - J_{5}^{\prime}$, is chosen such that the electric polarisation induced by the `high-temperature' configuration shown in figure~\ref{fig:ordering}(a) is $1000$ $\mu $C m$^{-2}$, then for the `low-temperature' configuration shown in figure~\ref{fig:ordering}(b) it equals $500$ $\mu $C m$^{-2}$.

\subsection{Phenomenological approach}

In this subsection we discuss the phenomenological description of spin states and the magnetoelectric coupling mechanism discussed above, which will clarify the similarities between $R$Mn$_2$O$_5$ and other multiferroic materials. The positions of Mn$^{3+}$ and Mn$^{4+}$ in the paramagnetic unit cell are shown in table \ref{tab:coordinates} and the eight vector order parameters can be found in table \ref{tab:orderparameters} (we use the notations of Bertaut \etal \cite{Bertaut1967}).

\begin{table}[tbh]
\caption{The coordinates of Mn ions, where $x \approx 0.41$, $y \approx 0.35$ and $z \approx 0.26$ (for  BiMn$_2$O$_5$).}
\centering
\begin{tabular}{lcl}
\hline
Mn$^{3+}$ & & Mn$^{4+}$\\
[0.5ex]
\hline
$\bi{r}_{1} = \left(x,y,1/2\right)$ & &
$\bi{r}_{5} = \left(1/2,0,z\right)$ \\
[0.5ex]
$\bi{r}_{2} = \left(-x,-y,1/2\right)$ & & $\bi{r}_{6} = \left(1/2,0,-z\right)$ \\
[0.5ex]
$\bi{r}_{3} = \left(1/2-x,1/2+y,1/2\right)$ & &
$\bi{r}_{7} = \left(0,1/2,z\right)$ \\
[0.5ex]
$\bi{r}_{4} = \left(1/2+x,1/2-y,1/2\right)$ & &
$\bi{r}_{8} = \left(0,1/2,-z\right)$ \\
[1ex]
\hline
\end{tabular}
\label{tab:coordinates}
\end{table}

\begin{table}[tbh]
\caption{Magnetic order parameters.}
\centering
\begin{tabular}{ccc}
\hline
Mn$^{3+}$ & & Mn$^{4+}$\\
[0.5ex]
\hline
$\bi{F} = \bi{S}_{1}+\bi{S}_{2}+\bi{S}_{3}+\bi{S}_{4}$ & &
$\bi{F}^{\prime} = \bi{S}_{5}+\bi{S}_{6}+\bi{S}_{7}+\bi{S}_{8}$ \\
[0.5ex]
$\bi{C} = \bi{S}_{1}+\bi{S}_{2}-\bi{S}_{3}-\bi{S}_{4}$ & &
$\bi{C}^{\prime} = \bi{S}_{5}+\bi{S}_{6}-\bi{S}_{7}-\bi{S}_{8}$ \\
[0.5ex]
$\bi{G} = \bi{S}_{1}-\bi{S}_{2}+\bi{S}_{3}-\bi{S}_{4}$ & &
$\bi{G}^{\prime} = \bi{S}_{5}-\bi{S}_{6}+\bi{S}_{7}-\bi{S}_{8}$ \\
[0.5ex]
$\bi{A} = \bi{S}_{1}-\bi{S}_{2}-\bi{S}_{3}+\bi{S}_{4}$ & &
$\bi{A}^{\prime} = \bi{S}_{5}-\bi{S}_{6}-\bi{S}_{7}+\bi{S}_{8}$ \\
[1ex]
\hline
\end{tabular}
\label{tab:orderparameters}
\end{table}

\begin{table}[tbh]
\caption{Irreducible representations of the space group $Pbam$ for $\bi{Q} = \left(1/2,0,1/2\right)$. }
\centering
\begin{small}
\begin{tabular}{cccccccc}
\hline
& 2$_{x}$ & 2$_{y}$ & 2$_{z}$ & m$_{x}$ & m$_{y}$ & m$_{z}$ & I \\
[0.1ex]
\hline
 & & & & & & & \\
$\Gamma_{1}$ &
$\!\!\left(
\begin{array}{cc}
 \!\!\!\!0\!\! & \!\!1\!\!\\
\!\!\!\!-1\!\! & \!\!0\!\!
\end{array}
\right)\!\!$ &
$\!\!\left(
\begin{array}{cc}
 \!\!1\!\! & \!\!0\!\!\!\!\\
 \!\!0\!\! & \!\!-1\!\!\!\!
\end{array}
\right)\!\!$ &
$\!\!\left(
\begin{array}{cc}
\!\!\!\!0\!\! & \!\!-1\!\!\!\!\\
\!\!\!\!-1\!\! & \!\!0\!\!\!\!
\end{array}
\right)\!\!$ &
$\!\!\left(
\begin{array}{cc}
\!\!1\!\! & \!\!0\!\!\!\!\\
\!\!0\!\! & \!\!-1\!\!\!\!
\end{array}
\right)\!\!$ &
$\!\!\left(
\begin{array}{cc}
\!\!\!\!0\!\! & \!\!1\!\!\\
\!\!\!\!-1\!\! & \!\!0\!\!
\end{array}
\right)\!\!$ &
$\!\!\left(
\begin{array}{cc}
1 & 0\\
0 & 1
\end{array}
\right)\!\!$ &
$\!\!\left(
\begin{array}{cc}
\!\!\!\!0\!\! & \!\!-1\!\!\!\!\\
\!\!\!\!-1\!\! & \!\!0\!\!\!\!
\end{array}
\right)\!\!$ \\
[5ex]
$\Gamma_{2}$ &
$\!\!\left(
\begin{array}{cc}
 \!\!0\!\! & \!\!-1\!\!\!\!\\
\!\! 1\!\! & \!\!0\!\!\!\!
\end{array}
\right)\!\!$ &
$\!\!\left(
\begin{array}{cc}
 \!\!1\!\! & \!\!0\!\!\!\!\\
 \!\!0\!\! & \!\!-1\!\!\!\!
\end{array}
\right)\!\!$ &
$\!\!\left(
\begin{array}{cc}
0 & 1\\
1 & 0
\end{array}
\right)\!\!$ &
$\!\!\left(
\begin{array}{cc}
\!\!\!\!-1\!\! & \!\!0\!\!\\
\!\!\!\!0\!\! & \!\!1\!\!
\end{array}
\right)\!\!$ &
$\!\!\left(
\begin{array}{cc}
\!\!\!\!0\!\! & \!\!1\!\!\\
\!\!\!\!-1\!\! & \!\!0\!\!
\end{array}
\right)\!\!$ &
$\!\!\left(
\begin{array}{cc}
\!\!\!\!-1\!\! & \!\!0\!\!\!\!\\
\!\!\!\!0\!\! & \!\!-1\!\!\!\!
\end{array}
\right)\!\!$ &
$\!\!\left(
\begin{array}{cc}
\!\!\!\!0\!\! & \!\!-1\!\!\!\!\\
\!\!\!\!-1\!\! & \!\!0\!\!\!\!
\end{array}
\right)\!\!$ \\
 & & & & & & & \\
\hline
\end{tabular}
\end{small}
\label{tab:representations}
\end{table}

\begin{table}[tbh]
\caption{Basis vectors of the space group $Pbam$ for $\bi{Q} = \left(1/2,0,1/2\right)$.}
\centering
\begin{small}
\begin{tabular}{ccccccccccccc}
\hline
$\Gamma_{1}$ &
$\!\!\!\left(
\begin{array}{c}
 \!\!F_{x}\!\!\\
 \!\!C_{x}\!\!
\end{array}
\right)\!\!\!\!$ &
$\!\!\!\!\left(
\begin{array}{c}
\!\! C_{y} \!\!\\
\!\! F_{y} \!\!
\end{array}
\right)\!\!\!\!$ & &
$\!\!\!\!\left(
\begin{array}{c}
\!\!\!\!G_{x}\!\!\!\!\\
\!\!\!\!-A_{x}\!\!\!\!
\end{array}
\right)\!\!\!\!$ &
$\!\!\!\!\left(
\begin{array}{c}
\!\!\!\!-A_{y}\!\!\!\!\\
\!\!\!\!G_{y}\!\!\!\!
\end{array}
\right)\!\!\!\!$ & & & &
$\!\!\!\!\left(
\begin{array}{c}
\!\!C_{z}^{\prime}\!\!\\
\!\!F_{z}^{\prime}\!\!
\end{array}
\right)\!\!\!\!$ &
$\!\!\!\!\left(
\begin{array}{c}
\!\!\!\!G_{x}^{\prime}\!\!\!\!\\
\!\!\!\!-A_{x}^{\prime}\!\!\!\!
\end{array}
\right)\!\!\!\!$
&
$\!\!\!\!\left(
\begin{array}{c}
\!\!\!\!-A_{y}^{\prime}\!\!\!\!\\
\!\!\!\!G_{y}^{\prime}\!\!\!\!
\end{array}
\right)\!\!\!\!$ & \\
[5ex]
$\Gamma_{2}$ & & &
$\!\!\!\left(
\begin{array}{c}
 \!\!F_{z}\!\!\\
 \!\!C_{z}\!\!
\end{array}
\right)\!\!\!\!$ & & &
$\!\!\!\!\left(
\begin{array}{c}
\!\!\!\!G_{z}\!\!\!\!\\
\!\!\!\!-A_{z}\!\!\!\!
\end{array}
\right)\!\!\!\!$ &
$\!\!\!\!\left(
\begin{array}{c}
\!\!C_{x}^{\prime}\!\!\\
\!\!F_{x}^{\prime}\!\!
\end{array}
\right)\!\!\!\!$ &
$\!\!\!\!\left(
\begin{array}{c}
\!\!F_{y}^{\prime}\!\!\\
\!\!C_{y}^{\prime}\!\!
\end{array}
\right)\!\!\!\!$ & & & &
$\!\!\!\!\left(
\begin{array}{c}
\!\!\!\!G_{z}^{\prime}\!\!\!\!\\
\!\!\!\!-A_{z}^{\prime}\!\!\!\!
\end{array}
\right)\!\!\!\!$
\\
\hline
\end{tabular}
\end{small}
\label{tab:vectors}
\end{table}

For discussion of phenomenological description of the magnetoelectric coupling, we have chosen the relatively simple case of BiMn$_2$O$_5$, which shows the commensurate spin ordering with $\bi{Q} = \left(1/2,0,1/2\right)$. In this case the components of the order parameters belong to one of the two two-dimensional representations, $\Gamma_{1}$ or $\Gamma_{2}$, of the $Pbam$ group \cite{Bertaut1967,Munoz2002}. BiMn$_2$O$_5$ only shows the `collinear' state with the $a$-components of the Mn$^{3+}$ and Mn$^{4+}$ spins described, respectively, by the order parameters $F_{x} = -3.1 \mu_{\rm B}$ and $G_{x}^{\prime} = 2.4 \mu_{\rm B}$ with small $b$ components $C_{y} = - 0.8 \mu_{\rm B}$ and $A_{y}^{\prime} =  0.6 \mu_{\rm B}$ \cite{Munoz2002} corresponding to a small rotation between spins in neighbouring antiferromagnetic chains.

 Since $F_{x}$ is the part two-dimensional representation
$
\left(
\begin{array}{c}
F_{x} \\ C_{x}
\end{array}
\right)
\in \Gamma_{1},
$
the state described by the order parameter $C_{x}$ is another ground state of the system. These two states are related by inversion, which transforms $F_{x}$ into $-C_{x}$ and vice versa. It is easy to check that the magnetoelectric coupling of the form $-\lambda_{x} P_{y} \left(F_{x}^2 - C_{x}^{2}\right)$ is invariant upon all symmetry transformations of the paramagnetic phase, so that the order parameters $F_{x}$ and $C_{x}$ describe two ferroelectric states with opposite directions of electric polarisation. It also easy to check that the couplings $-\lambda_{y} P_{y} \left(F_{y}^2 - C_{y}^{2}\right)$ and $-\lambda_{z} P_{z} \left(F_{z}^2 - C_{z}^{2}\right)$ are also allowed by symmetry, which is a strong indication that the mechanism inducing electric polarisation in magnetically ordered state is invariant upon the global spin rotation and the coupling can be written in the form $\lambda P_{y} \left( \bi{F}^2 - \bi{C}^{2} \right)$ \cite{Kadomtseva}.

Due to the exchange coupling between the Mn$^{3+}$ and Mn$^{4+}$ ions, the order parameters $G_{x}^{\prime}$ and $F_{x}$ are strongly coupled. Since
$
\left(
\begin{array}{c}
G_{x}^{\prime} \\ -A_{x}^{\prime}
\end{array}
\right)
$
also belongs to $\Gamma_{1}$ representation, the coupling between the two spin subsystems is phenomenologically described by $-g \left(F_{x} G_{x}^{\prime} - C_{x} A_{x}^{\prime}\right)$ \cite{Kadomtseva}. Thus, more generally, the magnetoelectric coupling should be written in the form
\begin{equation}
\Phi_{\rm me} = - \lambda P_{y} \left(\eta_{1}^{2} - \eta_{2}^{2} \right),
\end{equation}
where
$
\left(
\begin{array}{c}
\eta_{1} \\ \eta_{2}
\end{array}
\right)
$
belongs to $\Gamma_{1}$ representation. This form of the third-order  magnetoelectric coupling was discussed previously in the context of the orthorombic manganites with the $E$-type magnetic ordering \cite{SergienkoPRL2006} and is typical for improper ferroelectrics \cite{Levanyuk}.

\subsection{Static and dynamic dielectric susceptibility}

The contribution of the coupled spin-lattice degrees of freedom to static dielectric susceptibility is given by
\begin{equation}
\chi _{1}^{-1}\approx \left( \chi _{1}^{\left( 0\right) }\right)
^{-1}-\frac{1}{V}\sum_{i,j}I_{i}A_{ij}^{-1}I_{j}+\frac{1}{2V}\sum_{i,j}J_{ij}^{\prime
\prime }\left( 0\right) \left( \bi{S}_{i}\bi{S}_{j}\right) ,
\label{eq:chistat}
\end{equation}
where
\begin{equation}
A_{ij}= J_{ij}\left( \bi{S}_{i}\bi{S}_{j}\right) +\delta
_{ij}\left[\sum_{\alpha =1,2}K_{i\alpha }\left( 2\left(
\bi{S}_{i}\hat{k}
_{i\alpha }\right) ^{2}-S_{i}^{2}\right) -  \sum_{k}J_{ik}\left( \bi{S}_{i}
\bi{S}_{k}\right) \right]
\label{eq:A}
\end{equation}

and
\begin{equation}
I_{i}=\sum_{j}J_{ij}^{\prime }\left[ \bi{S}_{i}\times \bi{S}_{j}
\right] _{c},  \label{eq:I}
\end{equation}
The second term in (\ref{eq:chistat}) is the spin contribution to the dielectric constant due to virtual excitations of magnons by electric field (this will become more apparent in the discussion of the dynamic susceptibility). 

The last term in (\ref{eq:chistat}) describes the shift of the phonon frequency due to a change of the spring constants in magnetically ordered states. 
This effect is known in condensed matter spectroscopy as spin-phonon coupling~\cite{Baltensperger,Sushkov-PRL2005}. 
Phenomenologically, this effect is described by the fourth-order magnetoelectric coupling of the type $P^2L^2$, where $L$ is a magnetic order parameter. 
In most cases, magnon and phonon branches coupled through this term experience `repulsion' and phonon hardens. 
However, in the magnetically frustrated compounds, this contribution to the dielectric constant can have either sign and can result in either hardening or softening of phonons in the magnetic phase (see, e. g., \cite{Wakamura1976} on CdCr$_2$S$_4$ spinel).

Equations of motion describing the coupled spin-lattice dynamics have the form,
\begin{equation}
\left\{
\begin{array}{ccc}
\ddot{P_{b}} & = & -\frac{\chi _{0}\omega _{0}^{2}}{V}\frac{\partial H }{\partial P_{b}}, \\
&  &  \\
{\dot \bi{S}}_{i} & = & \left[ \frac{\partial H }{\partial
\bi{S}_{i}}\times \bi{S}_{i}\right] ,
\end{array}
\right.
\label{eq:motion}
\end{equation}
where $\chi_{0} = \chi_{1}^{(0)} + \chi_{2}$ and $\omega _{0}$ is the bare frequency of the polar phonon. 
Omitting the fourth order coupling term and solving linearized equations of motion, we obtain the dynamic dielectric susceptibility:
\begin{equation}
\chi ^{-1}\left( \omega \right) \approx {\chi_{0}}^{-1}\left(1-\frac{\omega ^{2}}{\omega _{0}^{2}}\right) - \frac{1}{V}\sum_{i,j}I_{i}\left[ \left( BA-\omega ^{2}\right) ^{-1}B\right] _{ij}I_{j},
\label{chidyn}
\end{equation}
where
\begin{equation}
B_{ij} = J_{ij}+\frac{\delta_{ij}}{S_{i}^{2}} \sum_{\alpha =1,2}\left(K_{i\alpha }\left( \bi{S}_{i}{\hat k}_{i\alpha }\right)^{2}-K_{i3}S_{i}^{2}\right) -
\frac{\delta_{ij}}{S_{i}^{2}} \sum_{k}J_{ik}\left( \bi{S}_{i}\bi{S}_{k}\right) .
\label{eq:B}
\end{equation}

The second term in (\ref{chidyn}) describes the transfer of a part of electric dipole spectral weight from phonon to magnon frequencies, which turns magnons coupled to phonons into electromagnons.  If such an electromagnon has lower frequency than the phonon, the static dielectric constant $\varepsilon(0)=1+4\pi\chi(0)$ increases as a result of the coupling showing a step-like anomaly.
Furthermore, frequencies of the mixed spin-lattice excitations (poles of the dielectric susceptibility (\ref{chidyn})) are shifted down with respect to the `bare' magnon frequencies, found from
\begin{equation}
\det \left( BA-\omega^{2}\right) =0.
\label{eq:magfreq}
\end{equation}
Note that the electromagnon term in (\ref{chidyn}) disappears for collinear spin states, as $I_{i}$, defined by (\ref{eq:I}), is zero in this case. This can be understood as follows. Classically, magnons correspond to spin oscillations that are orthogonal to ordered spin vectors. The change of the scalar product of a pair of collinear spins is then proportional to the second power of the amplitude of the oscillations. Since the magnetoelectric coupling in our model originates solely from Heisenberg exchange and only involves scalar products of spins, the linear coupling of electric field to magnons is absent for collinear spins and the lowest-order process is the photo-excitation of a pair of magnons.

\begin{figure}[tbh]
\centering
\includegraphics[width=8cm]{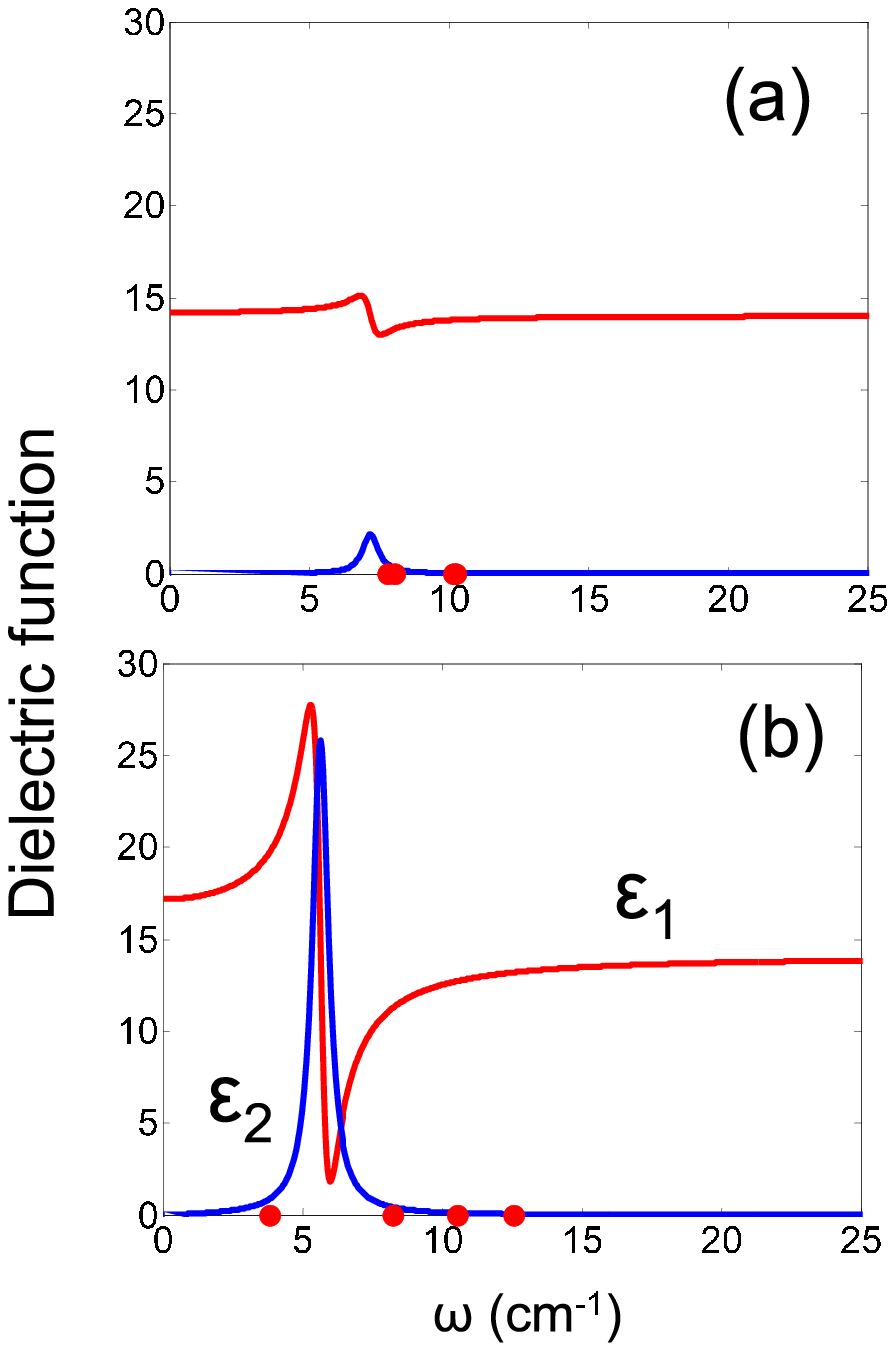} \vspace{0.0cm}
\caption{(Colour online) The model calculation results: Frequency dependence of the real ($\varepsilon_1$) and imaginary ($\varepsilon_2$) parts of the dielectric function (red and blue lines, respectively) for the `high-temperature' collinear state (panel a) and the `low-temperature' non-collinear state (panel b) shown, respectively, in panels (a) and (b) of figure~\ref{fig:ordering}. 
Red points are selected magnon frequencies of spins decoupled from the lattice found as the roots of (\ref{eq:magfreq}). 
A magnon at 11.8~cm$^{-1}$ couples to a polar phonon at 100~cm$^{-1}$ and becomes an electromagnon observable as the peak of the $\varepsilon_2$.}
\label{fig:modelspectra}
\end{figure}

In figures~\ref{fig:modelspectra}(a) and (b), we plot the real and imaginary parts of the dielectric function (red and blue lines, respectively) for the `high-temperature' and `low-temperature' states shown in, respectively, figures~\ref{fig:ordering}(a) and (b) (the imaginary part
was obtained by the shift $\omega \rightarrow $ $\omega +i\frac{\gamma }{2}$  with $\gamma =1$~K). 
As was explained above, the coupling of magnons to the electric component of light and significant electric dipole absorption at magnon frequencies is only present in the non-collinear `low-temperature' phase, in agreement with experiment. This result may seem somewhat counterintuitive: while the spontaneous electric polarisation induced by the Heisenberg exchange striction is maximum for the collinear state, the excitation of magnons by the electric component of light (electromagnons) requires non-collinear spins and is only observed below the spin re-orientation transition.

The red points in figure~\ref{fig:modelspectra} mark the bare frequencies of the softest magnons with zero wave vector, found from (\ref{eq:magfreq}). 
Since the magnetic unit cell in this calculation contains 16 mangetic ions, the total number of such magnons is also 16. 
However, only one of them is strongly coupled to the electric component of light and significantly contributes to the dielectric constant. 
This electromagnon corresponds to relative rotation of spins of the neighbouring antiferromagnetic chains, which gives rise to oscillations of the induced electric polarisation in the $b$ direction.  
The frequency of this uncoupled magnon for the non-collinear spin configuration shown in figure~\ref{fig:ordering}(b) is 11.8~cm$^{-1}$. 
As the position of the absorption peak in figure~\ref{fig:modelspectra}(b) is clearly lower that this magnon frequency, our parameters correspond to the strong magnetoelectric coupling case. The strong coupling is apparently necessary, if the large difference between dielectric constants of the `high'- and `low'-temperature
phases $\Delta \varepsilon ^{\prime }\left( 0\right) $ ($=3.25$ in our calculation) is associated solely with the absorption peak emerging in the non-collinear state. 

The results of our model calculations presented in figure~\ref{fig:modelspectra}(b) show that the symmetry properties of one magnon and the lattice allow an electromagnon in this system. 
Next, we discuss the strength of coupling.  
In our model, the first derivatives of the exchange integrals $J_{ij}$ are the coupling constants.
Both magnetically induced polarisation and electromagnon were calculated using the same set of $J^{\prime}$ values. 
Characteristic value of $J^{\prime}$ used in our calculation was $dJ_3/dy$ = 4~meV/$\AA$ which is much less than, for example, 40~meV/$\AA$ for ZnCr$_2$O$_4$ \cite{LeePRL}.

\section{Discussion}
\label{sec:discussion}

Our model calculations show that the observed values of the polarisation 
and electromagnon strength are obtained with realistic symmetric 
exchange constants. Also, the electromagnon coupling has been shown to 
be of Heisenberg type by the observed selection rules. These are strong 
arguments in favour of the Heisenberg exchange origin of the polarisation and electromagnons in 
$R$Mn$_2$O$_5$ family. However, in comparing the experimental and theoretical 
results, we should keep in mind that in the compounds of interest both 
symmetric and antisymmetric exchange mechanisms are operative, and they 
both may produce polarisation and electromagnons. Therefore, it is 
important to compare both the spontaneous polarisation and the 
electromagnon selection rules in both $R$MnO$_3$ and $R$Mn$_2$O$_5$ with 
these two mechanisms. 

In particular, the origin of ferroelectricity in $R$Mn$_2$O$_5$ remains controversial. 
An alternative to the magnetostriction scenario discussed here is electric polarisation induced by the $bc$ spiral, 
as observed in recent experiments (\cite{Noda2006,Vecchini2008,Kim2008}) 
via the inverse Dzyaloshinskii-Moriya mechanism of relativistic origin~\cite{Katsura2005,SergienkoDagotto,Mostovoy-spiral}. 
One natural question is whether the electromagnon study can clarify this controversy.  

Electromagnon excitations for the cycloidal spiral state were studied by 
Katsura \etal~\cite{Katsura-em}. According to their theory, the electric 
field of light that excites electromagnon has to be orthogonal to the 
direction of the electric polarisation induced by the spiral, which can 
be easily understood as follows. The spin spiral induces electric 
polarisation that lies in the spiral plane and is orthogonal to the 
propagation wave vector of the spiral \cite{Mostovoy-spiral}. Thus the 
$bc$-spiral in $R$Mn$_2$O$_5$ with the wave vector along the $c$ axis 
would induce polarisation along the $b$ axis, in agreement with 
experiment. An electric field applied in the direction perpendicular to 
the spiral plane will result in a small rotation of this plane, as a 
result of which the spontaneous polarisation vector acquires a component 
parallel to the applied field. An oscillating electric field orthogonal 
to the spiral plane will then excite oscillations of this plane, which 
is precisely the electromagnon of Katsura \etal~\cite{Katsura-em}. Since 
the polarisation lies in the spiral plane, the polarisation of light 
should be orthogonal to the spontaneous polarisation. 
Thus, to excite the 
oscillations of the $bc$ spiral would require $\bi{e} \| \bi{a}$, 
whereas in experiment electric field is parallel to the $b$ axis and the 
direction of electric polarisation. Indeed, this can be an argument 
against the spiral scenario of multiferroicity of $R$Mn$_2$O$_5$. 

However, the situation is complicated by the fact that even in 
$R$MnO$_3$ compounds, where the spiral origin of the 
magnetically-induced electric polarisation is well established, the 
selection rule $\bi{e} \perp \bi{P}$ is not obeyed. 
This was demonstrated by optical measurements in magnetic field. 
Magnetic field applied to DyMnO$_3$~\cite{Kida2008} and TbMnO$_3$~\cite{RolandoPhD} gives rise to the magnetic flop transition at 
which the $bc$-spiral is replaced by the $ab$-spiral. 
The spiral flop, however, has no effect on the polarisation of electromagnons, which are 
always observed for $\bi{e} \| \bi{a}$. 
Also in Eu$_{0.75}$Y$_{0.25}$MnO$_3$, where the spin spiral lies in the $ab$ 
plane already for zero magnetic field and the induced electric 
polarisation is parallel to $a$, the selection rule for electromagnons 
remains the same: ${\bi e} \| \bi{a}$ \cite{Valdes-EuY}. 
Therefore, the selection rules in both $R$MnO$_3$ and $R$Mn$_2$O$_5$ are tied to the 
lattice, not the spin plane. The selection rule for $R$MnO$_3$ appears 
to originate from the GdFeO$_3$ distortions of the perovskite structure 
of orthorombic manganites, which generates new types of magnetoelectric 
interactions and couples light to magnetic excitations that are 
different from the oscillations of the spiral plane \cite{Mostovoy113}. 
Furthermore, these distortions may also couple to the zone edge magnons 
and account for the broad peak observed in $R$MnO$_3$ at high frequencies. 
However, an alternative mechanism for the broad peak is the 
photoexcitation of bi-magnons and so the precise origin of this high 
frequency peak in $R$MnO$_3$ remains an open question. 

In any case, it appears that electromagnons in $R$MnO$_3$ are also induced 
by symmetric exchange even though the spontaneous polarisation is 
produced by the DM-type antisymmetric exchange mechanism. This is understood 
as a consequence of the vanishing of any static polarisation induced by 
symmetric exchange due to the symmetry of the lattice~\cite{SergienkoDagotto}. 
Another important question then is why the antisymmetric exchange does not 
produce electromagnon excitations? We suspect that the relativistic 
interactions are just too weak to produce observable signals. The 
observation of these weak DM induced electromagnons would help clarify 
the overall picture of of spin-lattice interactions in multiferroics. 

Consequently, while our experimental and theoretical results strongly 
suggest that the magnetoelectric coupling in $R$Mn$_2$O$_5$ is governed 
by Heisenberg exchange, we cannot safely rule out the spiral scenario 
for the spontaneous polarisation in these materials on the basis of optical 
data. However, the results of recent neutron scattering experiments~\cite{Kim2008} 
suggest rather that the spiral components appear as a consequence of the 
existence of spontaneous polarisation. 

\section{Summary}

We have presented experimental evidence and theoretical analysis that 
demonstrate that electromagnon excitations are present in both $R$Mn$_2$O$_5$ and $R$MnO$_3$ 
multiferroics. These electric dipole mixed 
magnon-phonon modes are observed in the far infrared and match excitations 
observed in inelastic neutron spectra in both classes of materials. The 
polarisation selection rules observed in the infrared experiments show 
that the electromagnon excitations are generated only by symmetric 
exchange in both classes of materials. The observed selection rules are 
tied to the lattice, not the spin plane, in contrast to the 
predictions of the antisymmetric exchange model. 

To theoretically account for electromagnons, we have considered third and 
fourth order coupling between the lattice and spins. The fourth order 
terms produce spin-phonon interactions that lead to shifts of the magnon 
and phonon frequencies near magnetic phase transitions observed in many 
magnetic materials. The third order coupling terms can produce mixed 
magnon-phonon excitations –-- the electromagnons --– only for non-collinear 
spin orders. The third order terms are also responsible for the 
spontaneous polarisation in multiferroics~\cite{Cheong2007}. We also 
developed a simple microscopic model of $R$Mn$_2$O$_5$ that explains the 
transition between collinear and non-collinear spin states, the 
magnetically-induced electric polarisation in the collinear state and 
the appearance of the electromagnon peak in the non-collinear state as 
well the polarisation of light for which this peak is observed. The 
mechanism of magnetoelectric coupling in this model relies entirely on 
isotropic Heisenberg exchange and magnetostrictive coupling of spins to 
a polar phonon mode. 

\ack This work was supported in part by the National Science Foundation MRSEC under Grant No. DMR-0520471. 
MM gratefully acknowledges the support by DFG (Mercator fellowship) and the hospitality of Cologne University. 
We acknowledge S.~Park, C.L.~Zhang and Y.J.~Choi for the growth and characterization of the single crystals.


\section*{References}

\begin{thebibliography}{99}

\bibitem{Smol-Chupis} Smolenskii G A and Chupis I E 1982 {\it Sov. Phys. Usp.} {\bf 25} 475

\bibitem{Fiebig2005} Fiebig M 2005 {\it J. Phys. D} {\bf 38} R123 

\bibitem{Prellier2005} Prellier W, Singh M P and Murugavel P 2005 \JPCM {\bf 17} R803 

\bibitem{Khomskii2006} Khomskii D I 2006 \JMMM {\bf 306} 1

\bibitem{Eerenstein2006} Eerenstein W, Mathur N D and Scott J F 2006 {\it Nature} {\bf 442} 759

\bibitem{Cheong2007} Cheong S-W and Mostovoy M 2007 {\it Nature Mater} {\bf 6} 13

\bibitem{Ramesh2007} Ramesh R and Spaldin N A 2007 {\it Nature Mater} {\bf 6} 21

\bibitem{Kimura-113} Kimura T, Goto T, Shintani H, Ishizaka K, Arima T and Tokura Y 2003 {\it Nature} {\bf 426} 55

\bibitem{Hur-nature} Hur N, Park S, Sharma P A, Ahn J S, Guha S and Cheong S-W 2004 {\it Nature} {\bf 429} 392

\bibitem{Lawes-vanadate} Lawes G, Harris A B, Kimura T, Rogado N, Cava R J, Aharony A, Entin-Wohlman O, Yildirim T, Kenzelmann M, Broholm C and Ramirez A P 2005 {\it Phys. Rev. Lett.} {\bf 95} 087205

\bibitem{Goto2004} Goto T, Kimura T, Lawes G, Ramirez A P, Tokura Y 2004 {\it Phys. Rev. Lett.} {\bf 92} 257201

\bibitem{Pimenov-Nature} Pimenov A, Mukhin A A, Ivanov V Yu, Travkin V D, Balbashov A M and Loidl A 2006 {\it Nature Phys.} {\bf 2} 97

\bibitem{Sushkov125} Sushkov A B, Vald\'{e}s Aguilar R, Park S, Cheong S-W and Drew H D 2007 {\it Phys. Rev. Lett.} {\bf 98} 027202

\bibitem{Katsura-em} Katsura H, Balatsky A V, and  Nagaosa N 2007 {\it Phys. Rev. Lett.} {\bf 98} 027203

\bibitem{Kimura07} Kimura T 2007 {\it Annu.\ Rev.\ Mater.\ Res.} {\bf 37} 387

\bibitem{Moriya} Moriya T 1960 {\it Phys. Rev.} {\bf 120} 91

\bibitem{Katsura2005} Katsura H, Nagaosa N and Balatsky A V 2005 {\it Phys. Rev. Lett.} {\bf 95} 057205

\bibitem{SergienkoDagotto} Sergienko I A and Dagotto E 2006 {\it Phys. Rev.} B {\bf 73} 094434

\bibitem{Jia} Jia C, Onoda S, Nagaosa N and Han J H 2007 {\it Phys. Rev.} B {\bf 76} 144424

\bibitem{Kadomtseva} Kadomtseva A M, Krotov S S, Popov  Yu F, Vorob'ev G P and Lukina M M 2005 {\it JETP} {\bf 100} 305

\bibitem{ChaponPRL2006} Chapon L C, Radaelli P G, Blake G R, Park S and Cheong S-W 2006 {\it Phys. Rev. Lett.} {\bf 96} 097601

\bibitem{SergienkoPRL2006} Sergienko I A, Sen C and Dagotto E 2006 {\it Phys.\ Rev.\ Lett.} {\bf 97} 227204

\bibitem{Baryachtar} Bary'achtar V G, L'vov V A and Jablonskii D A 1983 {\it JETP Letters} {\bf 37} 673 

\bibitem{Mostovoy-spiral} Mostovoy M 2006 {\it Phys. Rev. Lett.} {\bf 96} 067601

\bibitem{Levanyuk} Levanyuk A P and Sannikov D G 1974 {\it Sov. Phys. Usp.} {\bf 17} 199

\bibitem{Lorenzana1995} Lorenzana J and Sawatzky G A 1995 {\it Phys. Rev. Lett.} {\bf 74} 1867

\bibitem{Damascelli} Damascelli A, van der Marel D, M. Gr\"{u}ninger M, Presura C, Palstra T T M, Jegoudez J and Revcolevschi A 1998 {\it Phys. Rev. Lett.} {\bf 81} 918

\bibitem{Pimenov-PRB} Pimenov A, Rudolf T, Mayr F, Loidl A, Mukhin A A and Balbashov A M 2007 {\it Phys. Rev.} B {\bf 74} 100403(R)

\bibitem{Valdes-EuY} Vald\'{e}s Aguilar R, Sushkov A B, Zhang C L, Choi Y J, Cheong S-W and Drew H D 2007 {\it Phys. Rev.} B {\bf 76} 060404(R)

\bibitem{Pimenov-EuY} Pimenov A, Loidl A, Mukhin A A, Travkin V D, Ivanov V Yu and Balbashov A M 2008 {\it Phys. Rev.} B {\bf 77} 014438

\bibitem{Kida2008} Kida N, Ikebe Y, Takahashi Y, He J P,  Kaneko Y, Yamasaki Y, Shimano R, Arima T, Nagaosa N and Tokura Y 2007 arXiv:0711.2733

\bibitem{deSoussaPRB2008} de Sousa R and Moore E J 2008 {\it Phys. Lett.} B {\bf 77} 012406

\bibitem{deSousa} de Sousa R and Moore E J 2008 {\it Appl. Phys. Lett.} {\bf 92} 022514

\bibitem{FangCM} Fang C and Hu J 2008 {\it Europhys. Lett.} {\bf 82} 57005

\bibitem{Bertaut1967} Bertaut E F, Buisson G,  Quezel-Ambrunaz S and Quezel G 1967 {\it Solid State Commun.} {\bf 5} 25

\bibitem{Harris2008} Harris A B, Aharony A and Entin-Wohlman O 2008 {\it Phys. Rev. Lett.} {\bf 100} 217202 

\bibitem{Sanina1988} Sanina V A, Sapozhnikova L M, Golovenchits E I and Morozov N V 1988 {\it Sov. Phys. Solid State} {\bf 30} 1736

\bibitem{Mukhin2000} Mukhin A A, Travkin V D, Ivanov V Yu, Lebedev S P, Prokhorov A S and Kon K 2001 {\it Bulletin of the Lebedev Physics Institute} {\bf 4} 34

\bibitem{Golovenchits2003} Golovenchits E I and Sanina V A 2003 {\it JETP Lett.} {\bf 78} 88

\bibitem{reffit} Kuzmenko A B 2004 Reffit: Software to fit Optical Spectra (http://optics.unige.ch/alexey/reffit.html)

\bibitem{Kobayashi2004} Kobayashi S, Osawa T, Kimura H, Noda Y, Kagomiya I and Kohn K 2004 {\it J. Phys. Soc. Jpn.} {\bf 73} 1031

\bibitem{Sirenko-Ho} Sirenko A A, O Malley S M, Ahn K H, Park S, Carr G L and Cheong S-W 2007 arXiv:cond-mat/0703255v1

\bibitem{RolandoPhD} Vald\'{e}s Aguilar R 2008 PhD Thesis, University of Maryland

\bibitem{LeeAPS} Lee S-H, Kim J H, Chung J-H, Qiu Y, Kenzelmann M, Sato T J, Park S Y, Cheong S-W 2007 {\it Bull. Am. Phys. Soc.} {\bf 52} 330 

\bibitem{SenffPRL} Senff D, Link P, Hradil K, Hiess A, Regnault L P, Sidis Y, Aliouane N, Argyriou D N and Braden M 2007 {\it Phys. Rev. Lett.} {\bf 98} 137206

\bibitem{Noda2006} Noda Y, Kimura H, Kamada Y, Osawa T, Fukuda Y, Ishikawa Y, Kobayashi S, Wakabayashi Y, Sawa H, Ikeda N and Kohn K 2006 {\it Physica} B 385

\bibitem{Vecchini2008} Vecchini C, Chapon L C, Brown P J, Chatterji T, Park S, Cheong S-W and Radaelli P G 2008 {\it Phys. Rev.} B {\bf 77} 134434

\bibitem{Munoz2002} Mu\~{n}oz A, Alonso J A, Casais M T, Mart\'{i}nez-Lope M J, Mart\'{i}nez J L and  Fern\'{a}ndez-D\'{i}az M T 2002 {\it Phys. Rev.} B {\bf 65} 144423

\bibitem{Baltensperger} Baltensperger W and Helman J S 1968 {\it Helv. Phys. Acta} {\bf 41} 668

\bibitem{Sushkov-PRL2005} Sushkov A B, Tchernyshyov O and Ratcliff II W, Cheong S-W and Drew H D 2005 {\it Phys. Rev. Lett.} {\bf 94} 137202

\bibitem{Wakamura1976} Wakamura K and Arai T 1988 {\it J. Apl. Phys.} {\bf 63} 5824

\bibitem{LeePRL} Lee S-H, Broholm C, Kim T H, Ratcliff II W. and Cheong S-W 2000 {\it Phys. Rev. Lett.} {\bf 84} 3718

\bibitem{Kim2008} Kim T H \etal 
2008 arXiv:0803.1123v2

\bibitem{Mostovoy113} Mostovoy M \etal to be published

\end{thebibliography}
\end{document}